\documentclass[letterpaper,11pt]{article}

\usepackage{fullpage}
\usepackage{latexsym}
\usepackage{amsmath}
\usepackage{amssymb}
\usepackage{amsfonts}

\def\01{\{0,1\}}

\newcommand{\eps}{\varepsilon} 
 
\newcommand{\IP}{\mbox{IP}} 
\newcommand{\HAM}{\mbox{HAM}} 
\newcommand{\ket}[1]{|#1\rangle}
\newcommand{\bra}[1]{\langle#1|}
\newcommand{\ketbra}[2]{|#1\rangle\langle#2|}
\newcommand{\norm}[1]{\mbox{$\parallel{#1}\parallel$}}
\newcommand{\inp}[2]{\langle{#1}|{#2}\rangle} 
\newcommand{\Tr}{\mbox{\rm Tr}}
\newtheorem{definition}{Definition}
\newtheorem{theorem}{Theorem}
\newtheorem{lemma}{Lemma}

\newcommand{\f}[1]{\mbox{$#1$}}

\newenvironment{proof}
{\noindent {\bf Proof. }}
{{\hfill $\Box$}\\
 \smallskip}

\begin{document}

\title{Quantum Communication Cannot Simulate a Public Coin}
\author{Dmitry Gavinsky\thanks{University of Calgary.
gavinsky@cpsc.ucalgary.ca. Supported in part by Canada's NSERC.}
\and 
Julia Kempe\thanks{CNRS \&\ LRI, Univ.~de Paris-Sud, Orsay. Supported in part by ACI S\'ecurit\'e Informatique SI/03 511 and ACI-Cryptologie CR/02 20040 grants of the French
Research Ministry and the EU fifth framework project RESQ, IST-2001-37559.}
\and 
Ronald de Wolf\thanks{CWI, Amsterdam. rdewolf@cwi.nl. Supported in part by the EU fifth framework project RESQ, IST-2001-37559.}}
\date{}
\maketitle

\begin{abstract}
We study the simultaneous message passing model of communication complexity.
Building on the quantum fingerprinting protocol of Buhrman et al.,
Yao recently showed that a large class of efficient classical public-coin
protocols can be turned into efficient quantum protocols without public coin.
This raises the question whether this can be done always, i.e.~whether
quantum communication can always replace a public coin in the SMP model.
We answer this question in the negative, exhibiting a communication
problem where classical communication with public coin is exponentially
more efficient than quantum communication.
Together with a separation in the other direction due to Bar-Yossef et al.,
this shows that the quantum SMP model is incomparable with the classical
public-coin SMP model.

In addition we give a characterization of the power
of quantum fingerprinting by means of a connection to
geometrical tools from machine learning, a quadratic improvement
of Yao's simulation, and a nearly tight analysis of the Hamming distance
problem from Yao's paper.
\end{abstract}

\section{Introduction}

\subsection{Setting}

The area of communication complexity deals with the amount 
of communication required for solving computational problems 
with distributed input. 
This area is interesting in its own right, but also has many applications
to lower bounds on circuit size, data structures, etc.
The \emph{simultaneous message passing} (SMP) model involves three parties: 
Alice, Bob, and a referee.
Alice gets input $x$, Bob gets input $y$. They each
send one message to the referee, to enable him to compute something
depending on both $x$ and $y$, such as a Boolean function or some relational property.
The \emph{cost} or \emph{complexity} of a communication protocol 
is the length of the total communication for a worst-case input,
and the complexity of a problem is the cost of the best protocol. 

The SMP model is arguably the weakest setting of communication complexity
that is still interesting. Even this simple setting is 
not well understood. In the case of deterministic protocols, the optimal 
communication is determined by the number of distinct rows (and columns)
in the communication matrix, which is a simple property.
However, as soon as we add randomization to the model things become 
much more complicated.  For one, we can choose to either add \emph{public}
or \emph{private} coin flips. In more general communication models 
this difference affects the optimal communication by at most an additive
$O(\log n)$~\cite{newman:random}, but in the SMP model the difference can be huge.
For example, the equality function for $n$-bit strings requires about 
$\sqrt{n}$ bits of communication if the parties have only a private coin 
\cite{ambainis:3computer,newman&szegedy:1round,babai&kimmel:simultaneous}, 
but only constant communication with a public coin!
No simple characterization of public-coin
or private-coin communication complexity is known.\footnote{Kremer 
et~al.~\cite{knr:rand1round} claimed a characterization of public-coin 
complexity as the largest of the two one-way communication complexities, 
but Bar-Yossef et al.~\cite[Section~4]{bjks:itcc} 
exhibited a function where their characterization fails.}

The situation becomes more complicated still when we throw 
in \emph{quantum} communication. Buhrman et~al.~\cite{bcww:fp}
exhibited a no-coin quantum protocol for the equality function
with $O(\log n)$ qubits of communication. 
This is exponentially better
than classical private-coin protocols, but slightly worse than 
public-coin protocols.  
Roughly speaking, their quantum fingerprinting 
technique may be viewed as replacing the shared randomness by 
a quantum superposition.  This idea was generalized by Yao~\cite{yao:qfp},
who showed that every public-coin protocol with $c$-bit messages for 
a Boolean function can be simulated by a quantum fingerprinting protocol 
that uses $O(2^{4c}\log n)$ qubits of communication.
In particular, every $O(1)$-bit public-coin protocol can be simulated 
by an $O(\log n)$-qubit quantum protocol without public coin.   
Again, quantum superposition essentially replaces shared randomness in his construction. 

\subsection{Our results}

This raises the question whether something similar always holds in the SMP model:
can every efficient classical public-coin protocol be efficiently 
simulated by some quantum protocol without public coin? 
(Here we do not allow the quantum Alice and Bob to start with an entangled
state, since this could simulate a public coin for free.)
Since the appearance of Yao's paper, quite a number of people have tried to address this.
Our main result is a negative answer to this question. 
Suppose Alice receives input $x\in\01^n$ and Bob receives
$y,s\in\01^n$ with the property that $s$ has Hamming weight $n/2$.
The referee is required to output a triple $(i,x_i,y_i)$ for 
an $i$ satisfying $s_i=1$, but he is also allowed to output ``don't know'' 
with some small probability $\eps$.
We prove that public-coin protocols can solve this task with $O(\log n)$ 
bits of communication, while every quantum protocol needs $\Omega(\sqrt{n})$ 
qubits of communication.  This shows for the first time that the resource 
of public coin flips cannot efficiently be traded for quantum communication.
Our proof of the quantum lower bound may be of wider interest for 
the way it treats the independence of Alice's and Bob's messages.

Yao's exponential simulation can be made to work for relations as well, 
and our quantum lower bound shows that it is essentially optimal,
since the required quantum communication is exponentially larger
than the classical public-coin complexity for our relational problem.  
We expect a similar gap to hold for (promise) Boolean 
functions as well. In Section~\ref{secfpcharacterization}
we describe a function for which we conjecture an exponential gap.  
So far, we have not been able to prove this.
Our separation complements a separation in the other direction:
Bar-Yossef et al.~\cite{bjk:q1way} exhibited a relational problem where
quantum SMP protocols are exponentially \emph{more efficient} than classical
SMP protocols even with a public coin (also in their case it is open 
whether there is a similar gap for a Boolean function).
Accordingly, the quantum SMP model is incomparable with the classical 
public-coin SMP model.

In addition to our main result, we address some other open problems from Yao's paper.
A quantum fingerprinting protocol is one based on estimating the
inner product between a fingerprint of Alice's input and a fingerprint 
of Bob's input. Both the protocols of Buhrman et~al.~\cite{bcww:fp} 
and of Yao~\cite{yao:qfp} are based on this technique.
Our main result says that no quantum protocol can efficiently solve
the above relational task, so in particular quantum fingerprinting 
protocols are not able to do this.  Still, quantum fingerprinting
is the only technique we know to get interesting quantum protocols 
in the SMP model, and it makes sense to study this technique in detail.  
In Section~\ref{secfpcharacterization} we analyze the power of 
quantum fingerprinting, and tightly characterize it in terms of the
optimal margin achievable by realizations of the computational problem
via an arrangement of homogeneous halfspaces.  
The latter notion is well studied in machine learning.
We also give a small improvement of Yao's simulation for Boolean functions
in Section~\ref{secimproveyao}, by shaving a factor of 2 from the exponent.
A similar improvement has independently been observed by 
Chakrabarti and Regev~\cite{chakrabarti&regev:hamd} and by Golinsky and Sen
\cite{golinsky&sen:qfp}. The latter also extend Yao's definition of convex width.
Finally, we give nearly tight bounds on the quantum SMP 
complexity of the Hamming distance problem, which was the main application 
of Yao's simulation in his paper.

\section{Preliminaries}

Communication complexity, in particular the simultaneous message passing
model discussed here, is very intuitive so we will not provide formal
definitions. Instead we refer to Kushilevitz and Nisan~\cite{kushilevitz&nisan:cc}
for classical communication complexity and to the 
surveys~\cite{klauck:qccsurvey,buhrman:qccsurvey,wolf:qccsurvey} for the quantum variant.
We use $R_\eps^{\parallel}(P)$, $R_\eps^{\parallel,pub}(P)$, $Q_\eps^{\parallel}(P)$
to denote, respectively, the classical private-coin, classical public-coin, and quantum 
private-coin communication complexities with worst-case error probability $\eps$ 
for a distributed problem $P$. When the subscript is omitted, we take $\eps=1/3$.

The essentials needed for this paper are quantum states and their
measurement.  First, an $m$-qubit \emph{pure state} is a superposition 
$\ket{\phi}=\sum_{z\in\01^m}\alpha_z\ket{z}$ over all classical $m$-bit states.
The $\alpha_z$'s are complex numbers called \emph{amplitudes}, and 
$\sum_z|\alpha_z|^2=1$. Hence a pure state $\ket{\phi}$ is a unit vector in $\mathbb{C}^{2^m}$.
Its complex conjugate (a row vector with entries conjugated) is denoted $\bra{\phi}$.
The inner product between $\ket{\phi}$ and $\ket{\psi}=\sum_z\beta_z\ket{z}$ is
the dot product $\bra{\phi}\cdot\ket{\psi}=\inp{\phi}{\psi}=\sum_z\alpha_z^*\beta_z$.
The \emph{norm} of a vector $v$ is $\norm{v}=\sqrt{\inp{v}{v}}$.
Second, a \emph{mixed state} $\rho=\sum_i p_i\ketbra{\phi_i}{\phi_i}$
corresponds to a probability distribution over pure states, 
where $\ket{\phi_i}$ is given with probability $p_i$.
A $k$-outcome \emph{positive operator-valued measurement} (POVM) is given
by $k$ positive operators $E_1,\ldots,E_k$ with the property that $\sum_{i=1}^k E_i=I$.
When this POVM is applied to a mixed stated $\rho$, the probability
of the $i$-th outcome is given by the trace $\Tr(E_i\rho)$.
We refer to Nielsen and Chuang~\cite{nielsen&chuang:qc} for more details.

\section{Exponential Separation for a Relation}

In this section we prove our main result: we separate classical public-coin SMP
protocols from quantum protocols, by exhibiting a problem where the latter requires 
exponentially more communication.
Consider the following relational problem $P$:
\begin{quote}
Input: Alice receives input $x\in\01^n$, Bob receives $y,s\in\01^n$ with
$|s|=n/2$.\\ 
Output:
with probability $\geq 1-\eps$ output
$(i,x_i,y_i)$ s.t.~$s_i=1$, otherwise ``don't know''.
\end{quote}
This problem is easy for public coin-protocols, $R_\eps^{\parallel,pub}(P)=O(\log n\log(1/\eps))$:
Alice and Bob just send $(i,x_i)$ and $(i,y_i,s_i)$, respectively, 
to the referee for $\log(1/\eps)$ public random $i$'s. With probability $1-\eps$,
$s_i=1$ for at least one of those $i$'s and the referee outputs
the corresponding $(i,x_i,y_i)$. With probability $\eps$ he doesn't see an
$i$ for which $s_i=1$, in which case he outputs ``don't know''.
We assume here that the public coin is shared by Alice and Bob, not by the referee.
If we also allow the referee to view the coin, the classical public-coin 
complexity drops to $O(\log(1/\eps))$.
Below we show that we cannot trade the public coin for quantum communication: 
quantum SMP protocols need to communicate $\Omega(\sqrt{n})$ qubits to solve this task.

Consider a quantum protocol that solves $P$ 
and where Alice and Bob each send $m$-qubit messages.
Let $\alpha_x$ and $\beta_{ys}$ denote Alice's and Bob's messages produced in
response to the given $x$ and  \f{(y,s)}, respectively. 
These may be mixed states, reflecting private randomness. 

{\bf Alice's message cannot predict many $x_i$'s well.}\\
Define states
$\displaystyle\alpha_{i0}=\frac{1}{2^{n-1}}\sum_{x:x_i=0}\alpha_x$, \ 
$\displaystyle\alpha_{i1}=\frac{1}{2^{n-1}}\sum_{x:x_i=1}\alpha_x$, and
$\displaystyle\alpha=\frac{1}{2^n}\sum_{x}\alpha_x=\frac{1}{2}\alpha_{i0}+\frac{1}{2}\alpha_{i1}$.\\
Suppose we have a quantum measurement that does the following: given $i$, 
and either $\alpha_{i0}$ or $\alpha_{i1}$ (each with probability $1/2$), 
with probability $a_i$ it outputs $x_i$ (either 0 or 1),
and with probability $1-a_i$ it outputs ``don't know''.
We will use quantum information theory to show that 
most $a_i$'s must be fairly small. The ``average message'' $\alpha$
carries at least $a_i$ bits of information about $x_i$.
By Holevo's theorem~\cite{holevo}, the $m$-qubit $\alpha$ cannot 
contain more than $m$ bits of information, hence
$$
\sum_{i=1}^n a_i \leq m.
$$
By Markov's inequality, there exist $n/2$ $i$'s satisfying 
$a_i\leq 2m/n$.  In what follows, fix $s$ to the $n$-bit string 
corresponding to those $n/2$ $i$'s.

{\bf Bob's message cannot predict many $y_i$'s well.}\\
We similarly analyze Bob's average message $\beta_{ys}$ for 
our fixed $s$. Define $\beta_{i0}$ and $\beta_{i1}$ as before.
Let $b_i$ be the maximal probability with which we can output $y_i$, 
given $i$ and one of $\beta_{i0}$ or $\beta_{i1}$. Then 
$$
\sum_{i=1}^n b_i\leq m.
$$

{\bf Predictions of $x_i$ and $y_i$ are essentially independent.}\\
Now consider the referee. 
He does his measurement on the state $\alpha_x\otimes\beta_{ys}$, 
and with probability $p_i$ (probability taken over random $x$ and $y$), 
outputs $(i,x_i,y_i)$, and with probability $p_?\leq\eps$ he outputs ``don't know''. 
We now want to show that $p_i=O(a_i b_i)$. This follows from the next lemma,
which has the flavor of a ``direct product theorem'' from computational complexity.

\begin{lemma}\label{lemma:tensor}
A zero-error measurement for mixed states $\alpha_0$ and $\alpha_1$ is a 3-outcome 
measurement that outputs 0 or ? given $\alpha_0$, and outputs 1 or ? given $\alpha_1$.
Its success probability is the probability that it outputs 0 or 1 under
a uniform distribution over $\alpha_0$ and $\alpha_1$.
Let $a$ be the maximal success probability over all zero-error measurements
for $\alpha_0$ and $\alpha_1$ (so $1-a$ is the probability of output ?).
Define $b$ similarly for mixed states $\beta_0$ or $\beta_1$, 
and $p$ for 5-outcome zero-error measurements for $\alpha_0\otimes \beta_0 , \alpha_0\otimes \beta_1, 
\alpha_1\otimes \beta_0$, and $\alpha_1\otimes \beta_1$. Then $p \leq 4ab$.
\end{lemma}

\begin{proof}
Define $a^{(0)}=\max_{\cal M}\Pr[\mbox{measurement outputs 0 given }\alpha_0]$,
where ${\cal M}$ is the set of zero-error measurements for $\alpha_0$ and $\alpha_1$.
Similarly define $a^{(1)}$. Then it is easy to see that 
\begin{equation}\label{avmax}
\frac{1}{2}\max(a^{(0)},a^{(1)})\leq a \leq \frac{1}{2}(a^{(0)}+a^{(1)}).
\end{equation}
We can similarly define $b^{(0)}$ and $b^{(1)}$ for the measurement on $\beta_0$ or $\beta_1$,
and $p^{(00)},p^{(01)},p^{(10)},p^{(11)}$ for the 
measurement on $\alpha_0\otimes \beta_0 , \alpha_0\otimes \beta_1, 
\alpha_1\otimes \beta_0, \alpha_1\otimes \beta_1$. In particular
\begin{equation}\label{avmax2}
p \leq\frac{1}{4}(p^{(00)}+p^{(01)}+p^{(10)}+p^{(11)}).
\end{equation} 
Now we will show that $p^{(cd)}=a^{(c)}b^{(d)}$ for $c,d\in\01$. 
Without loss of generality assume $c=d=0$. 
Call ${\cal S}_0$ the support of $\alpha_0$ (i.e.~the span of the eigenstates 
with non-zero weight of $\alpha_0$), ${\cal S}_0^\perp$ its orthogonal complement,  
and similarly ${\cal T}_0$ and ${\cal T}^\perp_0$ for $\beta_0$. Note that 
because the measurement has zero error the support of the measurement operators 
corresponding to output ``0'' have to be in ${\cal S}_1^\perp$. Clearly the 
measurement that maximizes $a^{(0)}$ is the projection onto ${\cal S}_1^\perp$ and 
$a^{(0)}=\Tr(\alpha_0|_{{\cal S}_1^\perp})$. Similarly $b^{(0)}=\Tr(\beta_0|_{{\cal T}_1^\perp})$). 
To determine the measurement on $\alpha_0 \otimes \beta_0$ that gives 
the maximum $p^{(00)}$ note that each zero-error
measurement to determine $\alpha_0 \otimes \beta_0$ has to be 
in the intersection ${\cal I}=({\cal S}_0 \times {\cal T}_1)^\perp \cap ({\cal S}_1 \times {\cal T}_0)^\perp  \cap ({\cal S}_1 \times {\cal T}_1)^\perp $. 
Expanding $({\cal S}_0 \times {\cal T}_1)^\perp  = ({\cal S}^\perp_0 \times {\cal T}_1^\perp) \oplus ({\cal S}_0 \times {\cal T}_1^\perp ) \oplus ({\cal S}^\perp_0 \times {\cal T}_1)$ 
and similarly for the other terms we obtain ${\cal I} = {\cal S}^\perp_1 \times {\cal T}_1^\perp$, so the optimal measurement is a projection on ${\cal I}$, 
giving $p^{(00)}=\Tr[(\alpha_0 \otimes \beta_0)|_{({\cal S}^\perp_1 \times {\cal T}_1^\perp)}]=\Tr(\alpha_0|_{{\cal S}_1^\perp})\Tr(\beta_0|_{{\cal T}_1^\perp})=a^{(0)}b^{(0)}$.
The lemma now follows from (\ref{avmax}) and (\ref{avmax2}): 
$p \leq \frac{1}{4}(a^{(0)}b^{(0)}+a^{(0)}b^{(1)}+a^{(1)}b^{(0)}0+a^{(1)}b^{(1)}) 
\leq \frac{1}{4}(4\cdot 2a2b)=4ab.$
\end{proof}

{\bf Wrapping up the lower bound.}\\ 
Combining these building blocks gives $m=\Omega(\sqrt{n})$:
\begin{eqnarray*}
1-\eps & \leq & \Pr[\mbox{referee outputs $(i,x_i,y_i)$ for some $i$ with $s_i=1$}]\\
& \leq & \sum_{i:s_i=1} p_i
 \leq  \sum_{i:s_i=1} 4a_ib_i
 \leq  \frac{8m}{n}\sum_{i=1}^n b_i
 \leq  \frac{8m^2}{n}.
\end{eqnarray*}

{\bf A matching classical upper bound.}\\
This $\Omega(\sqrt{n})$ bound is tight, witness the following classical private-coin
protocol, inspired by~\cite{ambainis:3computer}. 
Alice and Bob view $n$-bit strings as $\sqrt{n}\times\sqrt{n}$
squares (we ignore rounding for simplicity). 
Alice sends the referee a randomly chosen row of $x$,
with its column-index, at the cost of $\sqrt{n}+\log\sqrt{n}$ bits.
Bob sends the referee a randomly chosen column of $y$, with its index,
and the same column of $s$ at the cost of $2\sqrt{n}+\log\sqrt{n}$ bits.  
Alice's row and Bob's column intersect in exactly one point $i\in[n]$, 
so for one (uniformly random) $i$ the referee now has $i,x_i,y_i,s_i$. 
With probability 1/2, $s_i=1$ and the referee is done.
Repeating this $\log(1/\eps)$ times costs $O(\sqrt{n}\log(1/\eps))$ bits 
and has error $\leq\eps$.

We summarize the exponential separation in our main theorem:

\begin{theorem}
For the problem $P$ we have $R^{\parallel,pub}(P)=\Theta(\log n)$ and
$R^{\parallel}(P),~Q^{\parallel}(P)=\Theta(\sqrt{n})$.
\end{theorem}

\emph{Remark:}  \f{R^{\parallel,pub}(P)=\Omega(\log n)} follows
from \f{Q^{\parallel}(P)=\Theta(\sqrt{n})} and from the extension of
Yao's exponential simulation to relations (as outlined at the end of 
Section~\ref{secimproveyao}). 
That simulation result is no longer true if the referee
sees the shared random string, since $R^{\parallel, pub}(P)$ drops to constant then.  

\section{Characterization of  Quantum Fingerprinting}\label{secfpcharacterization}

As mentioned, all nontrivial and nonclassical quantum SMP protocols known 
are based on a technique called quantum fingerprinting.
Here we will analyze the power of protocols that employ this technique, and 
show that it is closely related to a well studied notion from computational learning theory.
This addresses the 4th open problem Yao states in~\cite{yao:qfp}.
In particular, we will show that such quantum fingerprinting protocols 
cannot efficiently compute many Boolean functions for which there is
an efficient classical public-coin protocol.

Consider quantum protocols where Alice sends
a $q$-qubit state $\ket{\alpha_x}$, Bob sends a $q$-qubit state
$\ket{\beta_y}$, and the referee does the 2-outcome ``swap test''~\cite{bcww:fp}. 
This test outputs 0 with probability
$$
\frac{1}{2}+\frac{|\inp{\alpha_x}{\beta_y}|^2}{2}.
$$
They repeat this $r$ times in parallel, and the referee determines his output
based on the $r$ bits that are the outcomes of his $r$ swap tests.
We will call such protocols ``repeated fingerprinting protocols''.
The cost of the protocol is $2qr$.  For simplicity we assume
all amplitudes are real.
A quantum protocol of this form can only work efficiently if
we can ensure that $|\inp{\alpha_x}{\beta_y}|^2\leq \delta_0$ 
whenever $f(x,y)=0$ and $|\inp{\alpha_x}{\beta_y}|^2\geq \delta_1$ 
whenever $f(x,y)=1$.  Here $\delta_0<\delta_1$ should be reasonably 
far apart, otherwise $r$ would have to be too large to distinguish 
the two cases. A statistical argument shows that 
$r=\Theta(1/(\delta_1-\delta_0)^2)$ is necessary and sufficient.
We now define two geometrical concepts:

\begin{definition}
Let $f:{\cal D}\rightarrow\01$, with ${\cal D}\subseteq X\times Y$,
be a distributed, possibly partial, Boolean function. 
Consider an assignment of unit vectors $\alpha_x\in\mathbb{R}^d$, 
$\beta_y\in\mathbb{R}^d$ to all $x\in X$ and $y\in Y$.

This assignment is called a \emph{$(d,\delta_0,\delta_1)$-threshold embedding of $f$}
if $|\inp{\alpha_x}{\beta_y}|^2\leq\delta_0$ for all $(x,y)\in f^{-1}(0)$ 
and $|\inp{\alpha_x}{\beta_y}|^2\geq\delta_1$ for all $(x,y)\in f^{-1}(1)$.

The assignment is called a 
\emph{$d$-dimensional realization of $f$ with margin $\gamma>0$} if
$\inp{\alpha_x}{\beta_y}\geq\gamma$ for all $(x,y)\in f^{-1}(0)$ and
$\inp{\alpha_x}{\beta_y}\leq-\gamma$ for all $(x,y)\in f^{-1}(1)$.
\end{definition}

\begin{lemma}\label{lemembedding}
If there is a $(d,\delta_0,\delta_1)$-threshold embedding of $f$,
then there is a $(d^2+1)$-dimensional realization of $f$ with margin 
$\gamma=(\delta_1-\delta_0)/(2+\delta_1+\delta_0)$.

Conversely, if there is a $d$-dimensional realization of $f$ with margin $\gamma$,
then there is a $(d+1,\delta_0,\delta_1)$-threshold embedding of $f$
with $\delta_0=(1-\gamma)^2/4$ and $\delta_1=(1+\gamma)^2/4$.
\end{lemma}

\begin{proof}
Let $\alpha_x,\beta_y$ be the vectors in a $(d,\delta_0,\delta_1)$-threshold 
embedding of $f$.  For $a=(\delta_1+\delta_0)/(2+\delta_1+\delta_0)$,
define new vectors $\alpha'_x=(\sqrt{a},\sqrt{1-a}\cdot\alpha_x\otimes\alpha_x)$
and $\beta'_y=(\sqrt{a}, -\sqrt{1-a}\cdot\beta_y\otimes\beta_y)$.
These are unit vectors of dimension $d^2+1$.
Now
$$
\inp{\alpha'_x}{\beta'_y}=a-(1-a)|\inp{\alpha_x}{\beta_y}|^2.
$$
If $(x,y)\in f^{-1}(1)$, then $|\inp{\alpha_x}{\beta_y}|^2\geq\delta_1$
and hence 
$\inp{\alpha'_x}{\beta'_y}\leq a-(1-a)\delta_1=-\gamma.$
Similarly, $\inp{\alpha'_x}{\beta'_y}\geq\gamma$ for $(x,y)\in f^{-1}(0)$.

For the converse, let $\alpha_x,\beta_y$ be the vectors in a
$d$-dimensional realization of $f$ with margin $\gamma$.
Define new $(d+1)$-dimensional unit vectors 
$\alpha'_x=(1,\alpha_x)/\sqrt{2}$ and $\beta'_y=(1,-\beta_y)/\sqrt{2}$. Now
$$
|\inp{\alpha'_x}{\beta'_y}|^2=\frac{1}{4}\left(1-\inp{\alpha_x}{\beta_y}\right)^2.
$$
If $(x,y)\in f^{-1}(1)$, then $\inp{\alpha_x}{\beta_y}\leq-\gamma$
and hence 
$|\inp{\alpha'_x}{\beta'_y}|^2\geq\frac{1}{4}\left(1+\gamma\right)^2=\delta_1.$
A similar argument shows 
$|\inp{\alpha'_x}{\beta'_y}|^2\leq\frac{1}{4}\left(1-\gamma\right)^2=\delta_0$ 
for $(x,y)\in f^{-1}(0)$.
\end{proof}

Our notion of a ``threshold embedding'' is essentially 
Yao's~\cite[Section~6, question~4]{yao:qfp}, except that we square
the inner product instead of taking its absolute value,
since it's the square that appears in the swap test's probability.
Clearly, threshold embeddings and repeated fingerprinting protocols 
are essentially the same thing.
The notion of a ``realization'' is computational learning theory's 
notion of the realization of a concept class by an arrangement 
of homogeneous halfspaces.  The tradeoffs between dimension $d$ and 
margin $\gamma$ have been well studied~\cite{forster:probcc,fklmss:relations,fsss:optmargins}.
In particular, we can invoke a bound on the best achievable
margin of realizations due to Forster~\cite{forster:probcc}:

\begin{theorem}[Forster]
For $f:X\times Y\rightarrow\01$, define the $|X|\times|Y|$-matrix $M$ 
by $M_{xy}=(-1)^{f(x,y)}$.
Every realization of $f$ (irrespective of its dimension) has margin $\gamma$ at most
$\gamma\leq \norm{M}/\sqrt{|X|\cdot|Y|}$,
where $\norm{M}$ is the operator norm (largest singular value) of $M$.
In particular, if $f:\01^n\times\01^n\rightarrow\01$ is the inner product
function, then $\norm{M}=\sqrt{2^n}$ and hence $\gamma\leq 1/\sqrt{2^n}$.
\end{theorem}

Combining this with Lemma~\ref{lemembedding}, we see that 
a $(d,\delta_1,\delta_0)$-threshold embedding of the inner product 
function has $\delta_1-\delta_0=O(1/\sqrt{2^n})$.
In repeated fingerprinting protocols, we then need $r\approx 2^n$ 
different swap tests to enable the referee to reliably distinguish 
0-inputs from 1-inputs!
Now consider the following promise function,
using blocks of inner product functions ($\IP(x,y)=\sum_j x_jy_j \mod 2$):
\begin{quote}
Let $x=x^1\ldots x^m$ and $y=y^1\ldots y^m$, 
for $m=n/\log n$ and $x^i,y^i\in\01^{\log n}$.\\
Promise: there is a $b\in\01$ such that $\IP(x^i,y^i)=b$ for at least $2/3$ of the $i$.\\
Output: $f(x,y)=b$
\end{quote}
Clearly, $R^{\parallel,pub}(f)\leq 2\log n$: Alice and Bob just pick
a shared random $i$ and send $x^i$ and $y^i$ to the referee,
who outputs $\IP(x^i,y^i)$.
In contrast, $f$ cannot be computed efficiently by a repeated fingerprinting protocol:
for bitstrings $a,b\in\01^{\log n}$, set $x^i=a$ and $y^i=b$ for 
all $i\leq m/3$, and fix the other $2m/3$ $(x^i,y^i)$-blocks such that half 
of them have inner product 0 and the other half has inner product 1.  
Then $f(x,y)=\IP(a,b)$. Since any repeated fingerprinting protocol needs
about $n$ qubits to compute $\IP$ on $\log n$ qubits, the same lower
bound applies to our promise function $f$. This shows that Yao's exponential
simulation of public-coin protocols by repeated fingerprinting protocols
cannot be improved much without going outside the fingerprinting framework.
We actually conjecture that all quantum SMP protocols need $\Omega(\sqrt{n})$ qubits for this function.

In general, the preceding arguments show that we can't have an efficient
repeated fingerprinting protocol if $f$ cannot be realized 
with large margin. If the largest achievable margin is $\gamma$, 
the protocol will need $\Omega(1/\gamma^2)$ copies of 
$\ket{\alpha_x}$ and $\ket{\beta_y}$.
If all rows or all columns of the matrix $M$ are distinct, then we
know that $\ket{\alpha_x}$ and $\ket{\beta_y}$ need $\Omega(\log n)$ 
qubits, which gives an $\Omega(\log(n)/\gamma^2)$ lower bound.
We now show that this lower bound is essentially optimal.
Consider a realization of $f$ with maximal margin $\gamma$.
Its vectors may have very high dimension, but nearly the same 
margin can be achieved via a random projection to fairly low
dimension~\cite[Section~5, Lemma~2]{fklmss:relations}:

\begin{lemma}[FKLMSS]
A $d$-dimensional realization of $f$ with margin $\gamma$ can be converted 
into an $O((n/\gamma)^2)$-dimensional realization of $f$ with margin $\gamma/2$.
\end{lemma}

Using Lemma~\ref{lemembedding}, this gives us a 
$(d,\delta_1,\delta_0)$-threshold embedding of $f$ with
$d=O((n/\gamma)^2)$, $\delta_0=(1-\gamma/2)^2/4$ and 
$\delta_1=(1+\gamma/2)^2/4$. Note that $\delta_1-\delta_0=\gamma/2$.
This translates directly into a repeated fingerprinting protocol
with states $\ket{\alpha_x}$ and $\ket{\beta_y}$ of $d$ dimensions,
hence $O(\log(n/\gamma))$ qubits, and $r=O(1/\gamma^2)$.
For example, if $f$ is equality then $\gamma$
is constant, which implies an $O(\log n)$-qubit repeated fingerprinting 
protocol for equality (of course, we already had one with $r=1$).
In sum:

\begin{theorem}
For $f:X\times Y\rightarrow\01$, define the $|X|\times|Y|$-matrix $M$ 
by $M_{xy}=(-1)^{f(x,y)}$, and let $\gamma$ denote the largest margin
among all realizations of $M$. There exists a repeated fingerprinting 
protocol for $f$ that uses $r=O(1/\gamma^2)$ copies of 
$O(\log(n/\gamma))$-qubit states.
Conversely, every repeated fingerprinting protocol for $f$ needs
$\Omega(1/\gamma^2)$ copies of its $\ket{\alpha_x}$ and $\ket{\beta_y}$ 
states.
\end{theorem}

\section{An Improved Exponential Upper Bound}\label{secimproveyao}

We now describe a way to simulate a given classical public-coin protocol 
for a Boolean function $f$ by means of a quantum fingerprinting scheme. 
We assume for simplicity that all Alice's messages are $c$ bits
(our simulation actually gets better if one party's messages are shorter than the other's).  
While Yao's simulation from~\cite{yao:qfp} took $O(2^{4c}(c+\log n))$ qubits,
ours takes $O(2^{2c}(c+\log n))$ and is arguably a bit simpler as well. 
We may assume that the protocol uses only $\log n+O(1)$ bits of randomness,
i.e., $n'=O(n)$ possible random strings $r$, each equally likely,
and that the referee is deterministic~\cite{newman:random}.
Let $a_{rx}$ be the $c$-bit message that Alice sends with random string $r$ 
and input $x$, and similarly $b_{ry}$ for Bob's $c$-bit messages.
Let $R(a,b)\in\01$ denote the referee's output given messages $a$ and $b$.
Then the acceptance probability of the protocol on input $x$ and $y$ is
$$
P(x,y)=\frac{1}{n'}\sum_r R(a_{rx},b_{ry}).
$$
This differs from $f(x,y)$ by at most $\eps=1/3$.
We will derive a reasonably good threshold embedding from this.
For $A_{ry}=\{a:R(a,b_{ry})=1\}$, define
$$
\ket{\alpha_x}=\frac{1}{\sqrt{n'}}\sum_r\ket{r}\ket{a_{rx}}
\mbox{ \ and \ \ }
\ket{\beta_y}=\frac{1}{\sqrt{n'}}\sum_r\ket{r}\left(\frac{1}{\sqrt{2^{c}}}\sum_{a \in A_{ry}}\ket{a}+
\sqrt{\frac{2^{c}-|A_{ry}|}{2^c}}\ket{\mbox{dummy}}\right),
$$
where `dummy' is some extra dimension.
These states live in dimension $d=n'(2^c + 1)$
The crucial observation is the following bound on $\delta_0$ and $\delta_1$:
$$
\inp{\alpha_x}{\beta_y}=\frac{1}{n'}\sum_r \frac{1}{\sqrt{2^{c}}}R(a_{rx},b_{ry})=
\frac{1}{\sqrt{2^{c}}}P(x,y)=
\left\{
\begin{array}{ll}
\geq \frac{2}{3\cdot\sqrt{2^{c}}}& \mbox{ if } f(x,y)=1\\[2mm]
\leq \frac{1}{3\cdot\sqrt{2^{c}}}& \mbox{ if } f(x,y)=0
\end{array}
\right.
$$
The $O(2^{2c}(c+\log n))$ simulation follows immediately by repeated fingerprinting.
%

\bigskip

\emph{Remark:} Yao's simulation can be extended to work for relational problems
as well, with some extra overhead. Consider a classical public-coin protocol with
$c$-bit messages.  With ``$[\cdot]$'' denoting the truth value of a statement,
the inner product of the states 
$\sum_{r}\ket{r}\ket{[a_{rx}=a]}$ and $\sum_{r}\ket{r}\ket{2-[b_{ry}=b]}$
equals the probability that the referee receives messages $a,b$ in 
the classical public-coin protocol.  Hence given $2^{O(c)}$ copies of 
these states, for all $a,b$, we can simultaneously estimate
all these probabilities up to additive error $1/(100\cdot 2^{2c})$. 
This enables the quantum referee to simulate the classical referee's 
behavior with small error probability. Also zero-error properties 
can be preserved. Due to this observation,
$Q^{\parallel}(P)=\Omega(\sqrt{n})$ implies a tight classical 
bound $R^{\parallel,pub}(P)=\Omega(\log n)$. 


\section{The Hamming Distance Problem}

\subsection{Upper bounds}\label{sechamdupper}

As an example of his simulation, Yao~\cite{yao:qfp} 
considered the following Hamming distance problem:
\begin{quote}
$\HAM^{(d)}_n(x,y)=1$ iff the Hamming distance between $x$ and $y$ is $\Delta(x,y)\leq d$.
\end{quote}
For $d=0$, this is just the equality problem.
Yao showed $R^{\parallel,pub}(\HAM^{(d)}_n)=O(d^2)$ (actually, a better 
classical protocol may be derived from the earlier paper~\cite{fimnsw:multiparty}).
This implies $Q^\parallel(\HAM^{(d)}_n)=2^{O(d^2)}(d^2+\log n)$, hence
for constant $d$ the problem can be solved with $O(\log n)$ qubits.

The third open problem of his paper asks for better upper bounds
for this problem, both quantum and classical.
We give two different quantum SMP protocols that are more efficient than
Yao's. The first costs $O(d^2\log n)$ qubits, and is similar to
a protocol found independently by Chakrabarti and Regev~\cite{chakrabarti&regev:hamd}.
It derives a threshold embedding directly from Yao's classical construction 
in~\cite[Section~4]{yao:qfp}.  There, the length of the messages sent by 
the parties is $m=\Theta(d^2)$.
The referee accepts only if the Hamming distance between the messages
is below a certain threshold $t=\Theta(m)$.
Let $a_{rx}$ be Alice's message on random string $r$ and input $x$,
$a_{rxi}$ be the $i$-th bit of this message, and similarly for Bob.
Again we may assume $r$ ranges over a set of size $n'=O(n)$~\cite{newman:random}.
Yao shows that for uniformly random $r$ and $i$,
$\Pr[a_{rxi}=b_{ryi}]\leq t/m-\Theta(1/d)$ if $\Delta(x,y)\leq d$,
and $\Pr[a_{rxi}=b_{ryi}]\geq t/m+\Theta(1/d)$ if $\Delta(x,y)>d$.
Here $t/m=\Theta(1)$.
Now define the following $(\log(n')+2\log(d)+1)$-qubit states:
$$
\ket{\alpha_x}=\frac1{\sqrt{mn'}}
 \sum_r\ket{r}\sum_{1\le i\le m}\ket{i}\ket{a_{rxi}}
\mbox{ \ and \ \ }
\ket{\beta_y}=\frac1{\sqrt{mn'}}
 \sum_{r}\ket{r}\sum_{1\le i\le m}\ket{i}\ket{b_{ryi}}.
$$
Then
$$
\inp{\alpha_x}{\beta_y}=\frac1{mn'}
 \sum_{r}\sum_{1\le i\le m}\delta_{a_{rxi},b_{ryi}}=
 \Pr[a_{rxi}=b_{ryi}].
$$
This threshold embedding implies an $O(d^2\log n)$-qubit
repeated quantum fingerprinting protocol. 
%

Our second protocol is not of the repeated fingerprinting type,
though it uses fingerprints as a tool. 
It has complexity $O(d(\log n)^2)$. This is much cheaper than Yao's 
protocol for $d\gg \sqrt{\log\log n}$.
The idea is the following. For every $x$, there are only
$D=\sum_{i=0}^d{n\choose i}$ different $y$ for which the
function evaluates to 1 (the Hamming ball of radius $d$ 
around $x$). We let Alice and Bob send $O(\log D)$ copies of 
the fingerprints of $x$ and $y$, respectively, to the referee.
We assume these fingerprints are derived from a linear 
constant-distance error-correcting code $E:\01^n\rightarrow\01^m$, $m=O(n)$, as
$$
\ket{\phi_x}=\frac{1}{\sqrt{m}}\sum_{i=1}^m (-1)^{E(x)_i}\ket{i}.
$$ 
This enables the referee to test whether $x=y$ 
with error probability $\ll 1/D^2$, but also to make $D$ different
equality tests such that with high probability these 
tests all succeed simultaneously (test unitarily, copy 
the answer, and reverse to get almost the original state back).
The referee can change the $j$-th bit of the fingerprinted string,
i.e., map $\ket{\phi_x}\mapsto\ket{\phi_{x^j}}$ using
the unitary map $\ket{i}\mapsto(-1)^{E(e_j)_i}\ket{i}$,
since $E(x^j)=E(x)\oplus E(e_j)$ by linearity. 
Hence he can just try out all $D$ possibilities around $x$ and 
test whether $y$ equals any one of those $D$ strings. 
This suffices to compute $\HAM_n^{(d)}(x,y)$ at a cost of
$O(\log D\cdot \log n)=O(d(\log n)^2)$ qubits (and $O(\log n)$ for $d=0$).

The same idea gives a classical public-coin SMP
protocol that uses $O(\log D)=O(d\log n)$ bits of communication
(do the usual classical protocol for equality 
with error reduced to $\ll 1/D$).
For $d\gg\log n$, this is better than Yao's $O(d^2)$-bit protocol.

\subsection{Lower bounds}

Here we show that the above protocols are close to optimal,
by proving an $\Omega(d)$ lower bound on all 1-way quantum protocols 
for $\HAM^{(d)}_n$ (1-way protocols are 2-party protocols where Alice sends one
message to Bob, who computes the output).
Consider a protocol where Alice sends a $q$-qubit message 
$\rho_x$ to Bob, who then outputs 
$\HAM^{(d)}_n(x,y)$ with probability $\geq 2/3$. We will show that the messages
actually induce a \emph{quantum random access code} for the set of $d$-bit 
strings~\cite{nayak:qfa}. A random access code of a $d$-bit string $z$ allows
its user to recover each bit $z_i$ with probability at least $2/3$.
Invoking Nayak's $(1-H(2/3))d$ lower bound on the length of such encodings
then finishes off the proof (here $H(\cdot)$ is the binary entropy function). 

Suppose Alice has a $d$-bit string $z$.  She defines $x=z0^{n-d}$, 
and sends to Bob the Hamming weight $|z|$ and the quantum message $\rho_x$. 
The total communication is $\log d$ classical bits and $q$ qubits.
Bob looks at the $|z|$ that he received, and defines 
$y=e_i 1^{d+1-|z|} 0^{n-2d-1+|z|}$, where $e_i$ is the $d$-bit string 
that has a 1 only at position $i$.  Then Bob completes the protocol 
for $x$ and $y$.  We have:
\begin{quote}
$z_i=1$ iff $\Delta(x,y)=d$\\
$z_i=0$ iff $\Delta(x,y)=d+1$
\end{quote}
This gives Bob the bit $z_i$, with probability 2/3, for any $i\in[d]$ of his choice.
Accordingly, the pair $(|z|,\rho_x)$ forms a random access code for $z$, and Nayak's
bound implies $\log(d) +q \geq (1-H(2/3))d$.
The same lower bound, with a loss of a factor of 2, also applies 
to the model where Alice and Bob start out with an unlimited
amount of entanglement (such as shared EPR pairs)~\cite{klauck:qpcom}. 
Hence:

\begin{theorem}
Every quantum 1-way communication protocol for $\HAM^{(d)}_n$ needs
$\Omega(d)$ qubits of communication (even with unlimited prior entanglement).
\end{theorem}

This bound also holds for weaker models, such as all SMP models discussed
in this paper, including the public-coin ones.
With private coin, it is also easy to show an $\Omega(\log n)$ 
bound via a reduction from equality. 
We summarize our quantum SMP upper and lower bounds:

\begin{theorem}
$\Omega(d+\log n)\leq Q^\parallel(\HAM_n^{(d)})\leq O(d\log n\min(d,\log n))$.
\end{theorem}

\section{Summary and Open Problems}

In this paper we have shed some new light on the power 
of quantum simultaneous message passing protocols, in particular in 
comparison to classical public-coin protocols. Our main result is
\begin{itemize}
\item A relational problem where quantum SMP protocols
need exponentially more communication than classical public-coin ones.
This gives the first exponential separation between the two models,
and shows that the resource of public randomness cannot be traded 
efficiently for quantum communication.
\end{itemize}
In addition we proved
\begin{itemize}
\item A characterization of the optimal quantum fingerprinting scheme
in terms of the best margin achievable by arrangements of homogeneous halfspaces.
\item Every classical $c$-bit public-coin protocol can be simulated 
by a quantum SMP protocol using $O(2^{2c}(c+\log n))$ qubits of
communication, which is a quadratic improvement over~\cite{yao:qfp}.
\item We constructed better quantum SMP protocols for the Hamming distance
problem, and showed that these are close to optimal.
\end{itemize}
The main problem left open by this paper is to modify our separation to
work for a Boolean function, for instance the one from Section~\ref{secfpcharacterization}.
And, of course, we would like to find more quantum protocols
beating their classical counterparts in some way or other.

\subsection*{Acknowledgments}
We thank Oded Regev for fruitful discussions, in particular for his suggestion 
to aim at an exponential separation for a relation instead of a function.
We also thank Richard Cleve, Hartmut Klauck (the $O(d(\log n)^2)$ protocol from 
Section~\ref{sechamdupper} arose in discussions with him) and Pranab Sen.


\begin{thebibliography}{10}

\bibitem{ambainis:3computer}
A.~Ambainis.
\newblock Communication complexity in a 3-computer model.
\newblock {\em Algorithmica}, 16(3):298--301, 1996.

\bibitem{babai&kimmel:simultaneous}
L.~Babai and P.~G. Kimmel.
\newblock Randomized simultaneous messages: Solution of a problem of {Y}ao in
  communication complexity.
\newblock In {\em Proceedings of the 12th IEEE Conference on Computational
  Complexity}, pages 239--246, 1997.

\bibitem{bjk:q1way}
Z.~{Bar-Yossef}, T.~S. Jayram, and I.~Kerenidis.
\newblock Exponential separation of quantum and classical one-way communication
  complexity.
\newblock In {\em Proceedings of 36th ACM STOC}, pages 128--137, 2004.

\bibitem{bjks:itcc}
Z.~{Bar-Yossef}, T.~S. Jayram, R.~Kumar, and D.~Sivakumar.
\newblock Information theory methods in communication complexity.
\newblock In {\em Proceedings of 17th IEEE Conference on Computational
  Complexity}, pages 93--102, 2002.

\bibitem{buhrman:qccsurvey}
H.~Buhrman.
\newblock Quantum computing and communication complexity.
\newblock {\em EATCS Bulletin}, 70:131--141, February 2000.

\bibitem{bcww:fp}
H.~Buhrman, R.~Cleve, J.~Watrous, and R.~{de} Wolf.
\newblock Quantum fingerprinting.
\newblock {\em Physical Review Letters}, 87(16), September 26, 2001.
\newblock quant-ph/0102001.

\bibitem{chakrabarti&regev:hamd}
A.~Chakrabarti and O.~Regev.
\newblock Personal communication, January 2004.

\bibitem{fimnsw:multiparty}
J.~Feigenbaum, Y.~Ishai, T.~Malkin, K.~Nissim, M.~Strauss, and R.~Wright.
\newblock Secure multiparty computation of approximations.
\newblock In {\em Proceedings of 28th ICALP}, volume 2076 of {\em Lecture Notes
  in Computer Science}, pages 927--938. Springer, 2001.

\bibitem{forster:probcc}
J.~Forster.
\newblock A linear lower bound on the unbounded error probabilistic
  communication complexity.
\newblock In {\em Proceedings of 16th IEEE Conference on Computational
  Complexity}, pages 100--106, 2001.

\bibitem{fklmss:relations}
J.~Forster, M.~Krause, S.~Lokam, R.~Mubarakzjanov, N.~Schmitt, and H-U. Simon.
\newblock Relations between communication complexity, linear arrangements, and
  computational complexity.
\newblock In {\em Proceedings of 21th FSTTCS}, pages 171--182, 2001.

\bibitem{fsss:optmargins}
J.~Forster, N.~Schmitt, H-U. Simon, and T.~Suttorp.
\newblock Estimating the optimal margins of embeddings in {E}uclidean half
  spaces.
\newblock {\em Machine Learning}, 51:263--281, 2003.

\bibitem{golinsky&sen:qfp}
A.~Golinsky and P.~Sen.
\newblock A note on the power of quantum fingerprinting.
\newblock Unpublished manuscript, personal communication, December 2003.

\bibitem{holevo}
A.~S. Holevo.
\newblock Bounds for the quantity of information transmitted by a quantum
  communication channel.
\newblock {\em Problemy Peredachi Informatsii}, 9(3):3--11, 1973.
\newblock English translation in {\it Problems of Information Transmission},
  9:177--183, 1973.

\bibitem{klauck:qpcom}
H.~Klauck.
\newblock On quantum and probabilistic communication: {Las Vegas} and one-way
  protocols.
\newblock In {\em Proceedings of 32nd ACM STOC}, pages 644--651, 2000.

\bibitem{klauck:qccsurvey}
H.~Klauck.
\newblock Quantum communication complexity.
\newblock In {\em Proceedings of Workshop on Boolean Functions and Applications
  at 27th ICALP}, pages 241--252, 2000.
\newblock quant-ph/0005032.

\bibitem{knr:rand1round}
I.~Kremer, N.~Nisan, and D.~Ron.
\newblock On randomized one-round communication complexity.
\newblock {\em Computational Complexity}, 8(1):21--49, 1999.
\newblock Earlier version in STOC'95. Correction at {\tt
  http://www.eng.tau.ac.il/\~{ }danar/Public/KNR-fix.ps}.

\bibitem{kushilevitz&nisan:cc}
E.~Kushilevitz and N.~Nisan.
\newblock {\em Communication Complexity}.
\newblock Cambridge University Press, 1997.

\bibitem{nayak:qfa}
A.~Nayak.
\newblock Optimal lower bounds for quantum automata and random access codes.
\newblock In {\em Proceedings of 40th IEEE FOCS}, pages 369--376, 1999.
\newblock quant-ph/9904093.

\bibitem{newman:random}
I.~Newman.
\newblock Private vs.~common random bits in communication complexity.
\newblock {\em Information Processing Letters}, 39(2):67--71, 1991.

\bibitem{newman&szegedy:1round}
I.~Newman and M.~Szegedy.
\newblock Public vs.~private coin flips in one round communication games.
\newblock In {\em Proceedings of 28th ACM STOC}, pages 561--570, 1996.

\bibitem{nielsen&chuang:qc}
M.~A. Nielsen and I.~L. Chuang.
\newblock {\em Quantum Computation and Quantum Information}.
\newblock Cambridge University Press, 2000.

\bibitem{wolf:qccsurvey}
R.~{de} Wolf.
\newblock Quantum communication and complexity.
\newblock {\em Theoretical Computer Science}, 287(1):337--353, 2002.

\bibitem{yao:qfp}
A.~C-C. Yao.
\newblock On the power of quantum fingerprinting.
\newblock In {\em Proceedings of 35th ACM STOC}, pages 77--81, 2003.

\end{thebibliography}

\end{document}